\documentclass[prd,twocolumn,nofootinbib,reprint]{revtex4-1}
\usepackage{amsmath,amssymb,amsthm,ascmac}
\usepackage{graphicx,fancyhdr}

\theoremstyle{definition}

\begin{document}
\title{
Enhanced axion-photon coupling in GUT with hidden photon
}
\author{Ryuji Daido$^{1}$, Fuminobu Takahashi$^{1,2}$, Norimi Yokozaki$^{1}$}
\address{
\vspace{12pt}$^{1}$Department of Physics, Tohoku University,  
Sendai, Miyagi 980-8578, Japan \\
$^{2}$Kavli Institute for the Physics and Mathematics of the Universe (WPI),
University of Tokyo, Kashiwa 277--8583, Japan \vspace{5pt}}

\begin{abstract}
\begin{center}{\bf Abstract}\end{center}
We show that the axion coupling to photons can be enhanced in simple models with 
a single Peccei-Quinn field, if the gauge coupling unification is realized by a large kinetic mixing $\chi = {\cal O}(0.1)$ between 
hypercharge and unbroken hidden U(1)$_H$.  The key observation is that the U(1)$_H$ gauge coupling 
should be rather strong to induce such large kinetic mixing, leading to enhanced contributions of 
hidden matter fields to the electromagnetic anomaly. We find that the axion-photon coupling is enhanced by 
about a factor of 10\,-\,100 with respect to the GUT-axion models with $E/N = 8/3$.
\end{abstract}
\pacs{***}
\maketitle

\thispagestyle{fancy}
\rhead{ TU-1057 \\ IPMU 18-0026}
\renewcommand{\headrulewidth}{0pt}

\section{Introduction}\label{sec:intro}

The axion, $a$,  is a pseudo-Nambu-Goldstone boson associated with spontaneous 
breakdown of a global U(1)$_{\rm PQ}$ symmetry in the Peccei-Quinn (PQ) mechanism \cite{Peccei:1977hh,Peccei:1977ur,Weinberg:1977ma,Wilczek:1977pj}. 
Not only does the axion solves the strong CP problem, but it can also explain dark matter \cite{Preskill:1982cy,Abbott:1982af,Dine:1982ah},
which makes it one of the well-motivated physics beyond the standard model. 

The axion or axion-like particles
have been searched for by numerous experiments (see e.g. Refs.\,\cite{Graham:2015ouw,Irastorza:2018dyq} for recent reviews). 
Many of the on-going and planned experiments exploit the axion coupling to photons,
\begin{equation}
\mathcal{L}=\frac{g_{a\gamma\gamma}}{4}aF_{\mu\nu}\tilde{F}^{\mu\nu}, \label{eq:gagg}
\end{equation}
where $F_{\mu \nu}$ is the photon field strength, and $\tilde{F}^{\mu\nu}$ denotes its dual. 
Therefore,  the size of the axion-photon coupling $g_{a \gamma \gamma}$ for a given 
decay constant $f_a$ or axion mass $m_a$ is a very important input for such experiments.

The axion-photon coupling is induced by one-loop diagrams where those particles running in the
loop have both electric and U(1)$_{\rm PQ}$ charges. In fact, the coupling depends on detailed structure of the PQ sector \cite{Sikivie:1986gq}. 
As an extreme example, one can consider
a clockwork (or aligned) QCD axion model~\cite{Higaki:2015jag,Higaki:2016yqk,Higaki:2016jjh}, 
which include multiple PQ fields with a clockwork structure~\cite{Kim:2004rp,Choi:2014rja,Choi:2015fiu,Kaplan:2015fuy,Giudice:2016yja}.
In this case, the axion can have a coupling to photons~\cite{Farina:2016tgd} 
or hidden photons \cite{Higaki:2016yqk}, which is exponentially larger
than gluons. 
However, in simple models 
with a single PQ field, there is a preferred range of $g_{a \gamma \gamma}$ for a given $f_a$~\cite{Kim:1998va,DiLuzio:2016sbl}.
We show two representative cases as black dash-dotted lines in Fig.~\ref{fig:agg}. 
In this paper we limit ourselves to such simple models with a single PQ scalar whose expectation value is equal to the decay constant (up to the domain wall number, $N_{DW}$).

Another important motivation for physics beyond the standard model is grand unified theories (GUTs). 
In a non-supersymmetric GUT, however, the unification scale tends to be too low to satisfy the proton decay constraint,
and moreover, the gauge couplings seem to fail unify at a single scale. 
One of the remedies for the gauge coupling unification is to add a massless hidden photon with a large kinetic mixing
with hypercharge, U(1)$_Y$~\cite{Redondo:2008zf}. According to the recent analysis using the two-loop renormalization group (RG) equations~\cite{Daido:2016kez},
the unification scale is shown to be at $10^{16.5}$\,GeV and the required kinetic mixing is $\chi(m_Z) \approx 0.37$.
Interestingly, the unification with a hidden photon is rather robust against adding visible or hidden matters~\cite{Takahashi:2016iph,Daido:2016kez}.
%
This finding enables us to incorporate the axion into
the framework in a consistent manner.
%

%


In this paper we study the axion coupling to photons in a GUT scenario where a massless hidden photon 
has a large kinetic mixing with hypercharge. 
Since the kinetic mixing between U(1)$_Y$ and U(1)$_H$
is induced by one-loop diagrams with bi-charged particles running in the loop, it requires either many such particles and/or rather strong hidden U(1)$_H$ gauge coupling~\cite{Takahashi:2016iph, Daido:2016kez}. As we shall see,
the large kinetic mixing and strong U(1)$_H$ gauge coupling enhance the electromagnetic anomaly, and
the axion coupling to photons can be enhanced even in a simple model with a single PQ field.
Such enhancement is advantageous for the axion search experiments utilizing the the axion photon
coupling.

The rest of this paper is organized as follows. In Sec.~II, we estimate the axion-photon coupling 
in the presence of hidden photons with large kinetic mixing and hidden matter fields charged
under U(1)$_{PQ}$.
In Sec.~III, we show that the axion-photon coupling can be indeed enhanced in the GUT scenario with U(1)$_H$. 
Finally, the last section is devoted to discussion and conclusions.


\section{Axion coupling to photons and kinetic mixing }\label{sec:axion}
In this section, we show how the axion-photon coupling is modified in the presence of unbroken hidden gauge symmetry, U(1)$_H$, which mixes with U(1)$_Y$.
First, let us review the standard case without U(1)$_H$. 
We assume that the global U(1)$_{\rm PQ}$ symmetry is spontaneously broken by a single complex scalar field $\phi$
whose PQ charge is set to be unity.
The potential of $\phi$ is given by
\begin{eqnarray}
V = \lambda_{PQ} \left( |\phi|^2 - \frac{v_{PQ}^2}{2} \right)^2,
\end{eqnarray}
with $\lambda_{PQ}>0$, and $\phi$ contains the axion in its phase component:
\begin{eqnarray}
\phi = \frac{v_{PQ} + \rho(x)}{\sqrt{2}} \exp\left(i \frac{a(x)}{v_{PQ}}\right).
\end{eqnarray}
The radial field $\rho(x)$ has a large mass around $v_{PQ}$, and it is irrelevant 
for our discussion. 


The global U(1)$_{\rm PQ}$ symmetry is assumed to be explicitly broken by the QCD anomaly.
To this end one introduces  heavy PQ fermions, $\psi_L^{(i)}$ and $\psi_R^{(i)}$, which couple to $\phi$ as
\begin{equation}
\sum_i \phi \,\bar \psi_L^{(i)} \psi_R^{(i)} + {\rm h.c.}.
\end{equation}
Here and in what follows we assign PQ charges $1$ and  $0$ on $\psi_L^{(i)}$ and $\psi_R^{(i)}$, respectively. 
We assume that the PQ fermions include PQ quarks charged under SU(3)$_C$. 
Through one-loop diagrams with the PQ quarks running in the loop,
the axion couples to the QCD anomaly,
\begin{equation}
\frac{g_s^2}{32\pi^2 f_a}a \,G^a_{\mu\nu}\tilde{G}^{a\mu\nu},
\end{equation}
where $G^a_{\mu\nu}$ is the gluon field strength, $\tilde{G}^{a\mu\nu}$ is its dual, and $f_a=v_{PQ}/N_{DW}$ 
is the decay constant of the QCD axion. In the above example, $N_{DW}$ is equal to the number of the 
heavy PQ quarks. 
The axion acquires a mass by the QCD instanton effects~\cite{diCortona:2015ldu},
\begin{equation}
m_a=5.70(7)\,\mu{\rm eV}\left(\frac{10^{12}\,{\rm GeV}}{f_a}\right).
\end{equation}
which is inversely proportional to $f_a$.

In general, the QCD axion also couples to photons through the electromagnetic anomaly and mixings with 
neutral mesons. The axion-photon coupling $g_{a\gamma\gamma}$ in Eq.(\ref{eq:gagg}) is given by~\cite{diCortona:2015ldu} 
\begin{equation}
g_{a\gamma\gamma} = \frac{\alpha_{\rm EM}}{2\pi f_a}\left(\frac{E}{N}-1.92(4)\right),\label{w/o_hidden}
\end{equation}
where $\alpha_{\rm EM}$ is the fine-structure constant, 
and $E$ and $N$ are the electromagnetic and color anomaly coefficients given by
\begin{align}
\label{EA}
E &=  \sum_i (Q^{(i)}_{\rm EM})^2  Q^{(i)}_{\rm PQ},\\
\label{CA}
N \delta_{ab} &=  \sum_i {\rm Tr} \lambda_a \lambda_b Q^{(i)}_{\rm {PQ}},
\end{align}
where $Q_{\rm EM}^{(i)}$ is the electric charge of $\psi^{(i)}$, $Q_{\rm PQ}^{(i)}$ the PQ charge of $\psi^{(i)}_L$,
$\lambda_a$  the generators for the PQ quarks under SU(3). For the fundamental representation of SU(3)$_C$,
we have $N = \frac{1}{2} \sum Q_{\rm PQ}$.
The ratio of the electromagnetic and color anomaly coefficients, $E/N$, 
is equal to $8/3$ if the PQ fermions form complete multiplets under SU(5)$_{\rm GUT}$,
and equal to $0$ if the PQ fermions do not  carry any electric charges. 
So the axion-photon coupling is determined by the gauge coupling constant and the anomaly coefficient.

One can see from Eqs.~(\ref{EA}) and (\ref{CA}) that $E/N$ is enhanced if either $Q_{EM}$ or $Q_{PQ}$ is
large. As we shall see shortly, a large $Q_{EM}$ is induced in a GUT scenario with hidden photons.
On the other hand,  in the clockwork QCD axion models~\cite{Higaki:2015jag,Higaki:2016yqk,Higaki:2016jjh}, 
some of heavy PQ fermions have exponentially large $Q_{\rm PQ}$. Then, for appropriate charge assignment 
of the PQ fermions and  interactions with the PQ fields, the axion-photon coupling can be enhanced~\cite{Sikivie:1986gq,Farina:2016tgd}. Note that one needs asymmetry in the color and electric charge assignments in this case, 
since otherwise the enhancement due to exponentially large $Q_{PQ}$ would be canceled in the ratio $E/N$. 
This necessitates introduction of GUT-incomplete multiplets. In this sense, the above two scenarios are complementary
to each other.


Next, we consider the effect of U(1)$_H$ and its kinetic mixing with hypercharge.
In the original basis where the kinetic mixing is apparent, 
the kinetic terms of the hypercharge and hidden gauge bosons, ${A}_{Y\mu}'$ and ${A}_{H\mu}'$, are
\begin{equation}
\mathcal{L}_K=-\frac{1}{4}{F'}_Y^{\mu\nu}{F}_{Y\mu\nu}'-\frac{1}{4}{F'}_H^{\mu\nu}{F}_{H\mu\nu}'-\frac{\chi}{2}{F'}_Y^{\mu\nu}{F}_{H\mu\nu}',
\end{equation}
where ${F}_{\mu\nu}'$ and ${F}_{H\mu\nu}'$ are field strengths of U(1)$_Y$ and U(1)$_H$, respectively.  
Let us introduce a  PQ fermion $\psi (q_Y, q_H)$ charged under U(1)$_Y$ and U(1)$_H$.
The relevant part of the Lagrangian is 
\begin{eqnarray}
\mathcal{L}_\psi &=&- (k \phi \, \bar{\psi}_L  \psi_R+h.c.) \nonumber \\
&+& \bar{\psi}\gamma^\mu[q_Yg_Y'{A}_{Y\mu}'+q_Hg_H{A}_{H\mu}']\psi,
\end{eqnarray}
where $g_Y'$ and $g_H$ are gauge couplings of U(1)$_Y$ and U(1)$_H$ in the original basis.

One can make the gauge bosons canonically normalized by the following transformation:
 \begin{align}
 {A}_{Y\mu}' &=\frac{A_{Y\mu}}{\sqrt{1-\chi^2}},\quad {A}_{H\mu}'=A_{H\mu}-\frac{\chi}{\sqrt{1-\chi^2}}A_{Y\mu},\\
 \mathcal{L}_K&=-\frac{1}{4}F_Y^{\mu\nu}F_{Y\mu\nu}-\frac{1}{4}F_H^{\mu\nu} F_{H\mu\nu}.
 \end{align}
Then, in the canonical basis, the gauge interaction terms of $\psi$  are given by
\begin{align}
\bar{\psi}\gamma^\mu&(q_Yg_Y'{A}_{Y\mu}'+q_Hg_H{A}_{H\mu}')\psi= \nonumber \\ 
&+\bar{\psi}\gamma^\mu[(q_Y-q_{\rm eff})g_YA_{Y\mu}+q_Hg_H{A}_{H\mu}]\psi,
\end{align}
with
\begin{equation}
g_Y=\frac{g_Y'}{\sqrt{1-\chi^2}},\quad q_{\rm eff}=q_H\frac{\chi}{\sqrt{1-\chi^2}}\frac{g_H}{g_Y}.
\end{equation}
One can see that hypercharge gauge coupling $g_Y$ in the canonical basis is larger than $g_Y'$ in the original
basis,  while $g_H$ remains unchanged under the transformation. 
It is important to note that the hidden charged particle acquires an effective hypercharge $q_{\rm eff}$ 
in the canonical basis even if $q_Y=0$. In this section we set $q_Y=0$ for simplicity.
(In the next section we also consider a case with $q_Y \ne 0$.)

Due to the effective hypercharge, the hidden charged particle also contributes 
to the electromagnetic anomaly. Its contribution $\Delta E$ is
\begin{eqnarray}
\Delta E =  \frac{ q_H^2\chi^2}{1-\chi^2}\frac{g_H^2}{g_Y^2}
\end{eqnarray}
 where the right-hand side is evaluated at the mass of $\psi$, $m_{\psi}=k v_{PQ} /\sqrt{2}$.\footnote{
In fact, $\Delta E$ is invariant at a scale below $m_{\psi}$ under the RG running 
at the one-loop level.
 }
Note that $g_{a\gamma\gamma}$ can be significantly enhanced for $\chi=\mathcal{O}(0.1)$ and $q_H g_H=\mathcal{O}(1)$. For instance, we obtain $\Delta E \approx 23$ for $\chi=0.44$, $q_H g_H=4.4$ and $g_Y=0.45$, where those values are motivated by the GUT scenario with $m_{\psi}=10^{16}$\,GeV.

Lastly let us mention other interactions induced by the hidden charged PQ fermions,
\begin{eqnarray}
\frac{g_{a\gamma\gamma'}}{4}aF_{\mu\nu}\tilde{F}^{\mu\nu}_H, \ 
\frac{g_{a\gamma'\gamma'}}{4}aF_{H \mu\nu}\tilde{F}^{\mu\nu}_H,
\end{eqnarray}
with
\begin{eqnarray}
g_{a\gamma\gamma'} &=& \frac{\sqrt{\alpha_{\rm EM} \alpha_{H}}}{\pi f_a} \left(-
\frac{1}{N} \frac{\chi}{\sqrt{1-\chi^2}} \frac{q_H^2 g_H}{g_Y}
\right) \nonumber \\
g_{a\gamma'\gamma'} &=& \frac{\alpha_{H}}{2\pi f_a} \left( \frac{1}{N}\, q_H^2 \right).
\end{eqnarray}
Here we have defined $\alpha_H=g_H^2/(4\pi)$. We will briefly discuss implications of the above
axion-photon-hidden photon and axion-hidden photon interactions in the last section. 

\section{Enhanced axion-photon coupling in GUT with U(1)$_H$}\label{sec:gut}

We have shown that $g_{a\gamma\gamma}$ is significantly enhanced if 
both $\chi$ and $g_H$ are large.
In fact, such large $\chi$ and $g_H$ are strongly favored by the GUT with U(1)$_H$,
as we shall see below. Here and in what follows we consider only complete multiplets under SU(5)$_{\rm GUT}$.

Firstly, the SM gauge couplings unify at around  $M_{\rm GUT} = 10^{16.5}$\,GeV with the kinetic mixing of $\chi(m_Z) \approx 0.37$  according to the analysis using the two-loop RGEs~\cite{Daido:2016kez}.\footnote{
In the analysis using one-loop RG equations~\cite{Redondo:2008zf,Takahashi:2016iph}, the unification scale was around $10^{17}$\,GeV with $\chi(m_Z)\approx 0.4$.
}
The unification is essentially determined only by $\chi(m_Z)$ and is insensitive to the size of $g_H$ 
nor the presence of visible and hidden matter fields at an intermediate scale~\cite{Takahashi:2016iph,Daido:2016kez}. 
Therefore a very large kinetic mixing is required by the GUT with U(1)$_H$.

Secondly, a rather large $g_H$ is required to induce such large kinetic mixing via loop diagrams 
involving bi-charged fields.
To see this, let us introduce $N_f$ bi-charged matter fields, $\Psi_{5_i}$, which transform
as {\bf 5} under SU(5)$_{\rm GUT}$ and has U(1)$_H$ charge of $q_H=-1$.
In order for $\Psi_{5_i}$ to induce a large kinetic mixing at the GUT scale, one needs to pick up
GUT-breaking effects because of the vanishing sum of hypercharge in the GUT complete multiplets. 
After the GUT breaking, $\Psi_{5_i}$ generically splits into SU(3)$_C$ triplet $\Psi_{D_i}$ and 
SU(2)$_L$ doublet $\Psi_{\bar L_i}$, respectively;
\begin{eqnarray}
-\mathcal{L} &\supset& \sum_{i=1}^{N_f} 
\left(M_5 \overline \Psi_{5_i}  \Psi_{5_i} + k \overline \Psi_{5_i} \left<\Sigma_{\rm 24} \right>  \Psi_{5_i} \right)  \nonumber \\
&=& \sum_{i=1}^{N_f}  \left(M_{D}
\overline \Psi_{D_i} \Psi_{D_i} + M_{L} \overline \Psi_{\bar L_i} \Psi_{\bar L_i}\right),
\end{eqnarray}
where $M_5 \sim M_{\rm GUT}$,  $\Sigma_{24}$ is a GUT breaking Higgs, 
 $g_{\rm GUT}$ is a coupling constant of SU(5)$_{\rm GUT}$, 
and $M_D$ and $M_L$ are masses of $\Psi_{D_i}$ and $\Psi_{\bar L_i}$, respectively.
Then, the induced kinetic mixing at one-loop level is estimated as
\begin{eqnarray}
\chi (M_{\rm GUT}) &\approx& 0.12 N_f \left(\frac{g_{\rm GUT}}{0.53}\right)  \nonumber \\
&\times&\left[\frac{g_H (M_{\rm GUT})}{4\pi}\right] 
\left[\frac{\ln (M_{D}/M_{L})}{\ln 4}\right].\label{eq:chi_gut}
\end{eqnarray}
%
We see that $N_f=\mathcal{O}(1)$ and $g_H(M_{\rm GUT}) \sim 4\pi$ induces the kinetic mixing of $\chi (M_{\rm GUT}) = \mathcal{O} (0.1)$ with a slight mass splitting between $M_D$ and $M_L$.\footnote{
With the mass splitting, we also have threshold corrections to the gauge coupling constants as
\begin{eqnarray}
\left[\Delta \alpha_3^{-1} -\Delta \alpha_2^{-1}\right]_{M_{\rm GUT}} &\simeq& \frac{N_f}{6\pi} \ln (M_{D}/M_{L}), \nonumber \\
\left[\Delta \alpha_1^{-1} -\Delta \alpha_2^{-1}\right]_{M_{\rm GUT}} &\simeq& \frac{2N_f}{15\pi} \ln (M_{D}/M_{L}), \nonumber \\
\left[\Delta \alpha_3^{-1} -\Delta \alpha_1^{-1}\right]_{M_{\rm GUT}} &\simeq& \frac{N_f}{5\pi} \ln (M_{D}/M_{L}). 
\end{eqnarray}
For a certain range of the parameters, the threshold corrections remain so small that 
their effect on the estimated $\chi$ is negligible, while the required $\chi = {\cal O}(0.1)$ is realized. 
}

\begin{figure}[!t]
\begin{center}
\includegraphics[scale=0.68]{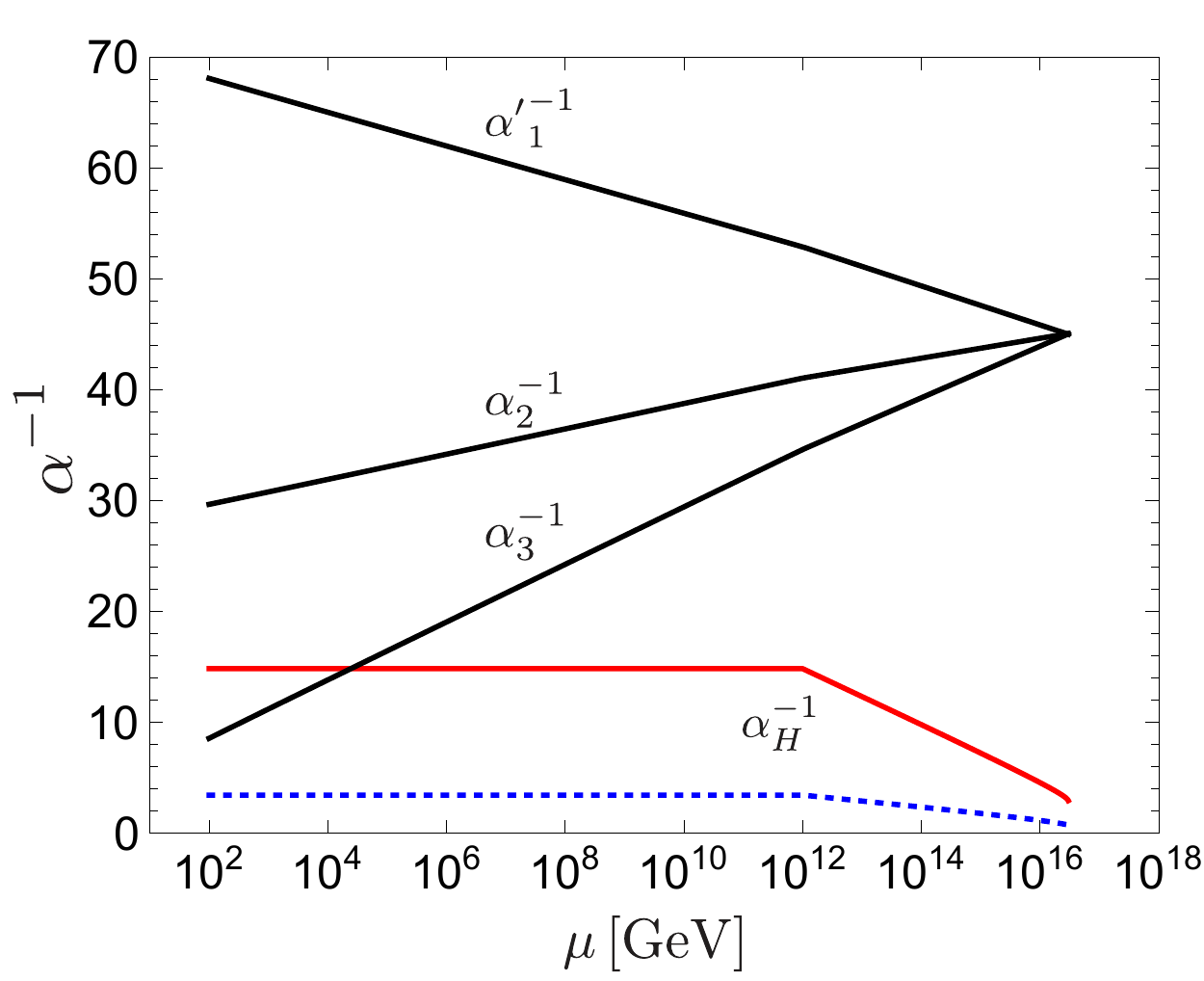}
\caption{The RG running of the gauge coupling constants.
We take $\chi(m_Z)=0.365$, $\alpha_s(m_Z)=0.1185$ and $m_t({\rm pole}) = 173.34$\,GeV.
}
\label{fig:unification}
\end{center}
\end{figure}

\begin{figure}[!t]
\begin{center}
\includegraphics[scale=0.68]{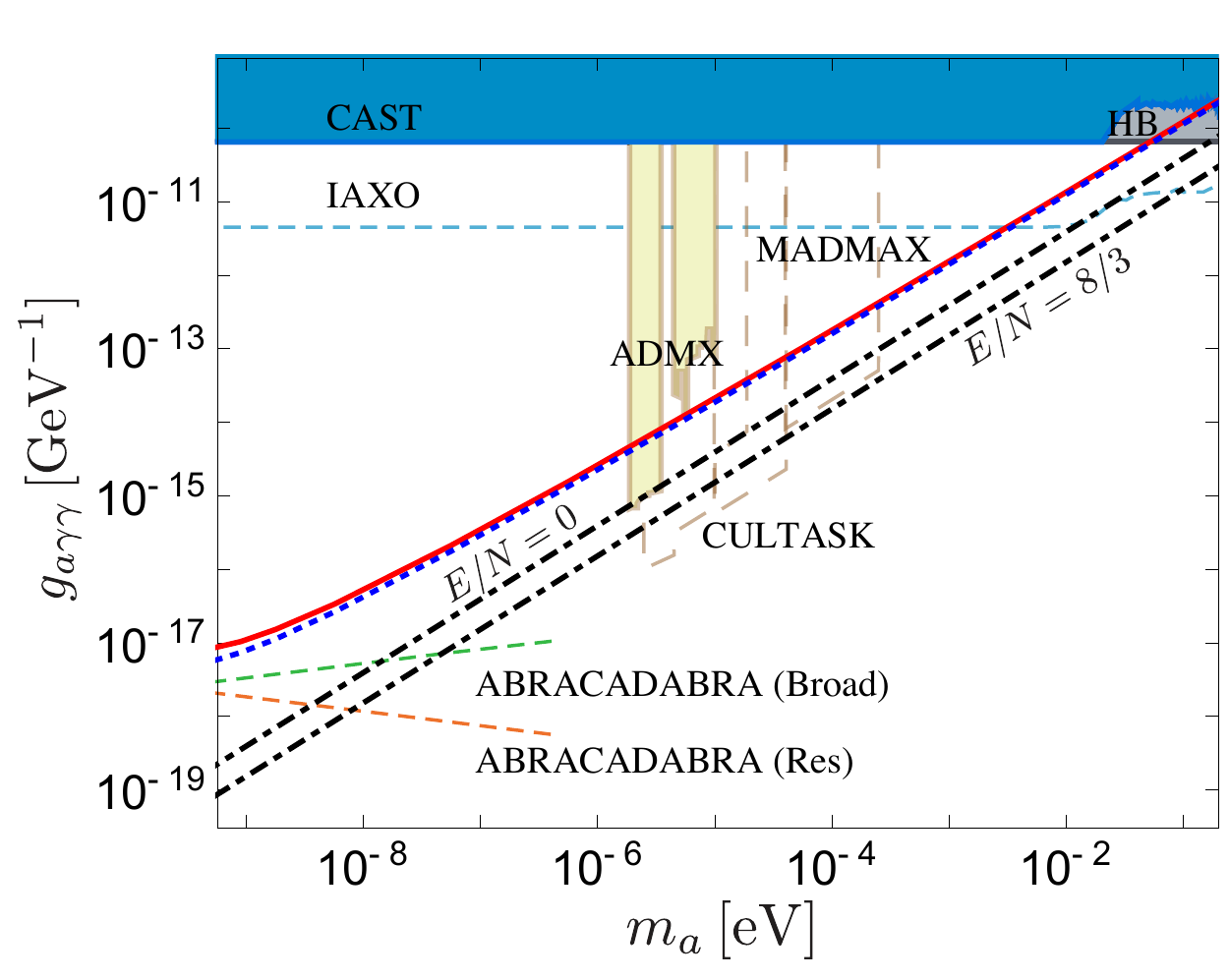}
\caption{The predicted axion-photon couplings as a function of the axion mass and experimental constraints.
The sensitivity reaches of future experiments are shown as dashed-lines.
}
\label{fig:agg}
\end{center}
\end{figure}

The RG runnings of the gauge coupling constants at the two-loop level are shown in Fig.~\ref{fig:unification}. We take $\chi(m_Z)=0.365$ and $\alpha_s(m_Z)=0.1185$. The black solid lines show the RG runnings of the SM gauge couplings, $\alpha_1$, $\alpha_2$ and $\alpha_3$ from top to bottom, 
where the normalization of U(1)$_Y$ is taken as $4 \pi \alpha_1=(5/3) {g_Y'}^2$.
We see that the SM gauge couplings unify at around $10^{16.5}$\,GeV. 
The blue dotted  and red solid lines show the running of the hidden gauge coupling in the following cases (i) and (ii),
respectively;
\begin{eqnarray}
&&{\rm Case\  (i)}: \mathcal{L} \supset - \left[\sqrt{2} \phi (\overline \psi_{5 L}  \psi_{5R} + \overline \psi_{H L} \psi_{H R}) + h.c.\right], \nonumber \\
&&{\rm Case\  (ii)}:\mathcal{L} \supset - \left[\sqrt{2} \phi \overline \psi_{5L}^{\rm b}  \psi_{5R}^{\rm b} + h.c.\right],
\end{eqnarray}
where $\psi_H$ is a hidden matter field with a charge of $q_H=1$, which is a SM gauge singlet; $\psi_{5}(0)$ and $\psi_{5}^{\rm b}(-1)$ transform as 
 {\bf 5} under SU(5)$_{\rm GUT}$ and their U(1)$_H$ charges are shown in the parentheses. (The superscript `b' means bi-charged.)
%
In the figure $v_{PQ} = f_a$ is taken to be $10^{12}$\,GeV.
In both cases, we take the largest possible $g_H$ avoiding the Landau pole below the GUT scale.
Note that the RG runnings of the SM gauge couplings are (almost) same in both case (i) and case (ii).

With the large $\chi$ and $g_H$ motivated by the GUT with U(1)$_H$, 
the axion-photon coupling $g_{a\gamma\gamma}$ is significantly enhanced. In Fig~\ref{fig:agg}, 
we show the predicted $g_{a\gamma\gamma}$ in the cases (i) and (ii), as well as 
 experimental/astrophysical constraints. As in the previous figure, we take $\chi(m_Z)=0.365$. The hidden gauge coupling, $g_H$, is taken as the largest possible value for a fixed $f_a$, avoiding the Landau pole below the GUT scale. The blue dotted (red solid) line corresponds to the case (i) (case (ii)), where the mass of the matter fields are set to be $f_a$. Interestingly, some part of the predicted region is already excluded by the ADMX experiment~\cite{Asztalos:2009yp,Carosi:2013rla}, and a large fraction of the region will be tested by future axion haloscopes such as ADMX~\cite{Asztalos:2003px}, CULTASK~\cite{Petrakou:2017epq}, MADMAX~\cite{TheMADMAXWorkingGroup:2016hpc}, ABRACADABRA~\cite{Kahn:2016aff}, whose sensitivity reaches are also shown in the figure.
The enhanced $g_{a \gamma \gamma}$ can be also reached by the next generation helioscopes. 
The sensitivity reach of IAXO~\cite{Irastorza:2011gs,Carosi:2013rla} is shown as blue-dashed line.

For comparison, the predicted $g_{a \gamma\gamma}$ in the usual case without U(1)$_H$ are also shown ($E/N=8/3$ and $E/N=0$). Here, $E/N=8/3$ corresponds to the case with $\mathcal{L} \supset - \sqrt{2} \phi (\overline \psi_{5L} \psi_{5R} + h.c.)$, which preserves the gauge coupling unification.
We see that $g_{a \gamma\gamma}$ in the case (ii) is enhanced by about a factor 10\,-\,100 for $f_a=10^{10}$-$10^{16}$\,GeV compared to the case of $E/N=8/3$.

Finally, we provide the approximate fitting formula for $g_{a \gamma \gamma}$ in the case (ii);
\begin{eqnarray}
\log_{10}(g_{a \gamma \gamma}/{\rm GeV}^{-1}) &\simeq& \left[
-8.9954 + 0.8862 x     \right. \nonumber \\
&-& \left.  0.0255 x^2 - 0.00285 x^3 \right],
\end{eqnarray}
where $x=\log_{10}(m_a/{\rm eV})$. The fitting formula is applicable in the range, $5.0 \times 10^{-10} {\rm\, eV} \leq m_a \leq 0.1 {\rm eV}$.

\section{Discussion and Conclusions}\label{sec:discuss}

We have shown that the axion coupling to photons can be enhanced in a simple model with 
a single Peccei-Quinn breaking field, if the gauge coupling unification is realized by a large kinetic mixing between 
hypercharge and unbroken hidden U(1)$_H$. 
The U(1)$_H$ gauge coupling should be rather strong to induce the large kinetic mixing (see Eq.(\ref{eq:chi_gut})). 
Consequently, matter fields charged under U(1)$_H$ significantly contribute to the electromagnetic anomaly: the axion-photon coupling is enhanced by about a factor 10\,-100 for $f_a=10^{10}$-$10^{16}$\,GeV, which can be tested in on-going and future experiments.

Let us comment on possible effects of $g_{a \gamma \gamma'}$ and $g_{a \gamma'\gamma'}$. 
For the kinetic mixing $\chi \approx 0.37$ required for the successful gauge coupling unification, 
these couplings are related to $g_{a \gamma \gamma}$ as 
$|g_{a\gamma\gamma'}|/g_{a\gamma\gamma}\simeq\,$5\,-\,6 and 
$g_{a\gamma'\gamma'}/g_{a\gamma\gamma}\simeq\,$8\,-\,9 for $f_a=10^{10}$-$10^{16}$\,GeV.
The coupling $g_{a \gamma \gamma'}$ contributes to extra stellar cooling through plasmon decay.
This is analogous to the stellar cooling due to neutrino dipole moment~\cite{Viaux:2013hca}. 
Based on our rough estimations, we obtain the constraint  $g_{a \gamma \gamma'} \lesssim 
\mathcal{O}(10^{-9})\,{\rm GeV}^{-1}$, which can be translated to $g_{a \gamma \gamma} \lesssim \mathcal{O}(10^{-10})\,{\rm GeV}^{-1}$.
This limit is not as stringent as the CAST bound, and so, this argument does not affect our results.
Recently, it was pointed out in Refs.~\cite{Agrawal:2017eqm,Kitajima:2017peg} that the other 
coupling, $g_{a \gamma'\gamma'}$, potentially suppresses the axion abundance 
via explosive production of hidden photons. According to the lattice calculations 
with $f_a = 10^{16}$\,GeV~\cite{Kitajima:2017peg}, the axion abundance is modified
for $g_{a \gamma'\gamma'}  \gtrsim 10/f_a$ and and the suppression factor is at most of order $10^2$
for $g_{a \gamma'\gamma'} \approx 200/f_a$. Such large coupling to hidden photons is easily realized
in the clockwork QCD axion models~\cite{Higaki:2015jag,Higaki:2016yqk,Higaki:2016jjh}, but not in simple models with
single PQ field. Therefore,  explosive hidden photon production does not take place in our set-up. 


\vspace{3pt}

\paragraph*{Acknowledgments:} This work is supported by JSPS Research Fellowships for Young Scientists (R.D.), 
Tohoku University Division for Interdisciplinary Advanced Research and Education (R.D.), JSPS KAKENHI Grant Numbers JP15H05889, JP15K21733, JP17H02875 (F.T and N.Y.) JP26247042 (F.T), JP17H02878 (F.T.),  JP17H05396 (N.Y.), and by World Premier International Research Center Initiative (WPI Initiative), MEXT, Japan (F.T.).

\end{document}